\documentclass[twocolumn, twocolappendix]{aastex631}
\usepackage{amsmath}

\newcommand{\Eorb}{$\Delta E_\mathrm{orb}$}

\shorttitle{Energy and Time in CE Hydrodynamics}
\shortauthors{Everson et al.}

\begin{document}

\title{The Energy Sharing Timescale in an Analytic Framework for Common Envelope Hydrodynamics}

\correspondingauthor{Rosa Wallace Everson}
\email{rosa@ucsc.edu}

\author[0000-0001-5256-3620]{Rosa Wallace Everson}
\affiliation{Department of Astronomy \& Astrophysics, University of California, Santa Cruz, CA 95064, USA}
\affiliation{Department of Physics and Astronomy, University of North Carolina at Chapel Hill, 120 E. Cameron Ave, Chapel Hill, NC 27599, USA}

\author[0000-0002-1417-8024]{Morgan MacLeod}
\affiliation{Institute for Theory \& Computation, Center for Astrophysics, Harvard \& Smithsonian, Cambridge, MA, 02138, USA}

\author[0000-0003-2558-3102]{Enrico Ramirez-Ruiz}
\affiliation{Department of Astronomy \& Astrophysics, University of California, Santa Cruz, CA 95064, USA}

\begin{abstract}
We propose a new predictive theory for the analysis of common envelope (CE) events which incorporates the effects of relevant hydrodynamical processes into a simple analytical framework. We introduce the ejection and dynamical parameters $\xi$ and $\beta$, which define if envelope ejection is energetically or hydrodynamically favorable, respectively, during CE inspiral. When combined, these parameters offer a detailed narrative of how inspiral begins, proceeds, and ends that is consistent with preliminary comparisons to 3D hydrodynamical models. This physically-motivated framework impacts predictions for CE outcomes, especially for systems that have energy excess, and offers promise as a potential alternative for the treatment of CE in binary population synthesis.
\end{abstract}

\section{Introduction} \label{sec:intro}

 A common envelope (CE) event occurs when one star in a binary system (hereafter, the primary) expands enough to engulf its partner (the secondary),  leading to the tightening of the binary's orbit \citep{Paczynski1976}. A CE episode can lead to the expulsion of the primary's stellar envelope, preserving the binary at a much smaller separation, or it can lead to a merger. Common envelope (CE) interactions are thought to be a critical aspect of the  formation of a broad range of close binary and multiple systems \citep{Iben1993,Taam2000,Taam2010,2013A&ARv..21...59I, Postnov2014}, such as the progenitors of gravitational wave sources \citep[e.g.,][]{2002ApJ...572..407B,Kruckow2016,Fragos2019,Renzo_2021}, unique products of interacting triples \citep[e.g.,][]{2016ComAC...3....6T,2020ApJ...901...44W,2024arXiv240408037R}, and Thorne-\.{Z}ytkow objects \citep{Tenley,2024ApJ...971..132E,2024arXiv240711680N}.

Extensive efforts have been undertaken to attempt to map pre-CE systems to their post-CE products \citep[e.g.,][]{vandenHeuvel1976,2010MNRAS.403..179D,Kruckow2016,2020PASA...37...38V,2021A&A...650A.107M}. Yet, CE evolution is complex, with key physical processes spanning a large range of length and time scales, requiring approximations in any mapping. The standard treatment is a decades-old energy formalism (see Section \ref{sec:energyformalism}) which parameterizes a simple energy criterion for envelope ejection via an efficiency term \citep{vandenHeuvel1976,Webbink1984,Livio1988,deKool1990,Iben1993}. It is well known that this prescription for envelope ejection is incomplete, if not problematic \citep[see, e.g.,][]{Soker1992,Soker2017, Clayton2017, Glanz2018,2020cee..book.....I}, folding many physical processes into a few parameters when perhaps additional criteria should be included.
In addition, there are known systems, such as close double white dwarfs and stellar-mass binary black holes, that the CE energy formalism cannot reproduce, though the use of convection or angular momentum as an alternative criterion has brought some relief \citep{2000A&A...360.1011N,2020MNRAS.497.1895W,2022ApJ...937L..42H,2022MNRAS.516.2189W,2023ApJ...944...87D}.

These issues shouldn't be surprising, as CE episodes are fundamentally hydrodynamic processes: the energy and angular momentum described in the aforementioned frameworks are all transported by waves, and in CE events, the speed of the secondary's plunge can often be faster than characteristic wave speeds in the local envelope. This means that the deposition of energy and angular momentum is not necessarily an equilibrium process, which is supported by numerous studies that apply 3D hydrodynamics codes to the CE problem \citep[for an in-depth review, see][]{2023LRCA....9....2R}. In fact, most global simulations of CE episodes do not realize the simple predictions of the energy formalism \citep[e.g.,][]{Passy2012,Ricker2012,Ohlmann2016,2016MNRAS.461..486K,2018MNRAS.477.2349I,Chamandy2019,2019MNRAS.490.2550I}.

In this Letter, we introduce a modified analytic framework that includes both energy budget and hydrodynamic communication time scale in the modeling of CE events. In Section \ref{sec:energyformalism} we review the energy formalism. In Section \ref{sec:analytical}, we present our modified model. In Section \ref{sec:results}, we illustrate the evolution of a selection of representative binary systems as interpreted via this mapping in 1D and 3D. We conclude after discussing implications and additional considerations in Section \ref{sec:discussion}.

\section{The Energy Formalism} \label{sec:energyformalism}
We begin by defining the standard CE energy formalism \citep{vandenHeuvel1976,Webbink1984,Livio1988,deKool1990,Iben1993}. Essentially, this formalism depends on the key assumption of energy conservation, which requires that the timescale of a CE event be much shorter than the thermal timescale of the envelope such that additional energy sources and sinks can be ignored. However, in cases in which the inspiral timescale approaches or exceeds the thermal timescale, energy conservation becomes less reliable as nuclear energy generation starts dominating the energy budget. Likewise, cases in which the inspiral timescale exceeds the envelope's convection timescale may be dominated by radiation losses \citep{2019MNRAS.485.4492W,2020MNRAS.497.1895W,2022MNRAS.516.2189W}. The result of a CE phase in this model  depends on the deposition of orbital energy into the binding energy of the envelope: if the energy change is at least equal to the binding energy, the envelope ejection is said to be successful. The orbital energy deposited due to orbital tightening is defined as
\begin{equation} \label{eqn:deltaEorb}
\Delta E_\mathrm{orb} (r) =  \frac{G M_\mathrm{enc} M_\mathrm{2}} 
{2 r} -\frac{G M_\mathrm{1} M_\mathrm{2}}{2 a_\mathrm{init}},
\end{equation}
in which $G$ is the gravitational constant, $M_\mathrm{1}$ is the total mass of the extended primary star, $M_\mathrm{2}$ is the mass of the embedded secondary, $a_\mathrm{init}$ is the initial separation of the primary and secondary, $r$ is the distance of the secondary from the primary's core after tightening, and $M_\mathrm{enc}$ is the enclosed mass of the primary within radius $r$ (i.e., $M_\mathrm{enc}=M(r)$).

It is understood that not all of the orbital energy will contribute to the ejection of the envelope  \citep{2018MNRAS.477.2349I,2019MNRAS.490.2550I}, and it is likely that other sources of energy may contribute \citep{Iben1993,2013A&ARv..21...59I}, e.g. recombination energy \citep[though this has been rigorously disputed, see][]{2017MNRAS.472.4361S,Grichener_2018}. Therefore, this quantity is modified by an efficiency term $\alpha$ that functions as a ``catch-all'' for energetic contributions and deficits during a CE phase: if considering only orbital energy deposition, then $\alpha \leq 1$, but if folding in contributions from other energy sources, $\alpha$ can increase the available energy budget by up to a factor of $\approx 5$ \citep{2020A&A...644A..60S}. It is this modified quantity that must be sufficient to unbind the envelope; this defines the so-called $\alpha$-formalism: 
\begin{equation} \label{eqn:alphaformalism}
    \alpha \Delta E_\mathrm{orb} \geq E_\mathrm{bind} ,
\end{equation}
in which $E_\mathrm{bind}$ is the total binding energy of the envelope. For simplicity, we adopt $\alpha=1$ and do not include additional energy source terms nor the specific internal energy of the gas \citep[though, see the discussion of][]{2015MNRAS.447.2181I,Soker2017,2017MNRAS.472.4361S,Grichener_2018,2018ApJ...863L..14S,2020MNRAS.494.5333R}, defining the total binding energy of the envelope as equivalent to the gravitational binding energy
\begin{equation} \label{eqn:Egrav}
E_\mathrm{grav} = - \int_{M_1}^{M_\mathrm{core}}  \frac{G M_\mathrm{enc}}{r} dm \equiv \frac{G M_1 M_\mathrm{env}}{\lambda R} ,
\end{equation}
in which $M_\mathrm{core}$ is the mass of the core of the primary, $M_\mathrm{env}$ is the total mass of the envelope of the primary ($M_1 - M_\mathrm{core}$), and $\lambda$ is a parameter included to account for differences in stellar structure amongst primaries \citep{deKool1990}. 

\section{Energy-Time Model} \label{sec:analytical}
\subsection{Energy: The Ejection Parameter\texorpdfstring{, $\xi$}{}} \label{subsec:ejectionparam}
The source of energy that we consider in the following is the orbital energy of the binary pair. It is transmitted to the CE through gravitational interaction with the secondary.
The characteristic gravitational length scale is the accretion radius as described by \citet{Hoyle1939} and \citet{Bondi1944} (hereafter HLA),
\begin{equation} \label{eqn:Ra}
R_\mathrm{a} = \frac{2 G M_\mathrm{2}}{v_{\infty}^2},
\end{equation}
in which $v_{\infty}$ is the velocity of the secondary relative to the envelope. 
Because the accretion radius describes the gravitational reach of the secondary deeper into the star, any gas within this distance contributes to the drag force on the secondary. If this material is removed, the drag experienced by the secondary decreases significantly. 

To approximate the binding energy that must be overcome not simply to eject the envelope outside mass coordinate $m$ but to remove the drag force and stop inspiral, we adjust Equation \ref{eqn:Egrav} to include the binding energy of material within an accretion radius of the secondary and present it in a local, integrated form such that
\begin{equation} \label{eqn:Egrav+}
\begin{split}
    E_{\mathrm{grav,SE}} (r) &:= E_{\mathrm{grav}} (r-R_\mathrm{a}) \\[1ex]
    &\, = - \int_{M_1}^{M(r-R_\mathrm{a})} \frac{G M_\mathrm{enc}}{r} dm .
\end{split}
\end{equation}
This adjustment to $E_\mathrm{grav,SE}$ is distinct in that it does not require the entire envelope to be ejected, but only the material exterior to $r-R_\mathrm{a}$, because this will be enough to effectively mitigate the drag force experienced by the secondary and halt its infall in envelope regions that contract in response to mass loss, or significantly slow its infall due to reduced drag in envelope regions that expand in response to mass loss.

To more easily parse where the energy criterion for ending the CE episode is satisfied, we introduce an ejection parameter, $\xi$, which specifies the total change in orbital energy at $r$, normalized by the gravitational binding energy of the envelope at $r-R_\mathrm{a}$, such that
\begin{equation} \label{eqn:xi}
    \xi (r) = 
    \frac{\Delta E_\mathrm{orb}(r)}{E_\mathrm{grav,SE}(r)} \equiv
    \frac{\Delta E_\mathrm{orb}(r)}{E_{\mathrm{grav}} (r-R_\mathrm{a})} .
\end{equation}
In the standard energy formalism, the energy criterion for envelope ejection at and above a given position is met where $\alpha \geq 1$; analogously, the energy criterion for envelope ejection sufficient to end inspiral (hereafter, sufficient ejection) at a given position is met where $\xi \geq 1$. We note that the assumption of energy conservation native to the standard CE energy formalism is maintained here, therefore nuclear energy generation and radiative losses are not considered.

\subsection{Time: The Dynamical Parameter\texorpdfstring{, $\beta$}{}} \label{subsec:dynamicalparam}

We argue that $\xi$ alone cannot predict the outcome of a CE interaction. Because energy transport does not occur instantaneously throughout the envelope, assessing the degree to which energy is shared can improve our model. Shocks launched by gravitational interaction between the secondary and the surrounding gas are the mechanism by which energy is shared and orbital decay forces are generated \citep[for a discussion of the analogous role of pressure waves in very low mass ratio systems, see][]{1993ApJ...417..347S}. 
Of interest here are how the orbital decay rate compares to the rate that shocks are able to diffuse through the envelope, redistributing energy and angular momentum.

We define a timescale for orbital decay due to CE drag as
\begin{equation} \label{eqn:tau_insp}
    \tau_\mathrm{insp} = - \frac{R_\mathrm{a}}{\dot{a}} ,
\end{equation}
in which $a$ is the semi-major axis of the secondary's orbit. 
The inspiral timescale quantifies how long it takes for the secondary to plunge on the order of its gravitational reach. We note that if $\tau_\mathrm{insp}$ is similar to or less than the orbital period throughout inspiral, then in principle the secondary is continuously plunging into undisturbed envelope. 

We estimate the timescale for energy transport through a shell of envelope material using wave speeds relevant to CE inspiral and the local environment at $r$ as
\begin{equation} \label{eqn:tau_SH}
    \frac{\pi r}{v_\mathrm{Kep}} < \tau_\mathrm{SH} < \frac{\pi r}{c_\mathrm{s}} ,
\end{equation}
in which the maximum velocity for the shock induced by the secondary ($v_\mathrm{Kep}$) and the sound speed of the gas at $r$ ($c_\mathrm{s}$) provide lower and upper limits\footnote{These limits apply to lateral energy sharing; however the convective velocity may in some cases be comparable to these wave speeds, leading to energy losses through radial transport to the photosphere. This would have no impact on $\beta$, but would effectively reduce $\xi$.}. This is a measure of the traversal time of the heat source in the envelope.

We suggest it is the way that these two timescales relate that reveals how inspiral proceeds. We introduce the dynamical parameter, $\beta$, as the ratio of interest:
\begin{equation} \label{eqn:beta}
    \beta = \frac{\tau_\mathrm{insp}}{\tau_\mathrm{SH}} .
\end{equation}
This is a measure of the timescale of shock-heating in the envelope relative to the timescale of the secondary's plunge. A high value of $\beta$ indicates that energy deposited by shocks is shared effectively throughout the CE. The ongoing infall of the secondary through the envelope is ``regulated'' by the energy deposition in that the addition of energy leads to adjustment of the surrounding envelope and modulation of the drag forces \citep[see, e.g.,][]{2019MNRAS.490.3727C}.  A low value of $\beta$ indicates that the infall is so rapid compared to energy sharing that any coupling between the energy deposited and the envelope local to the secondary is weak. 

\subsection{Energy-Time \texorpdfstring{$\xi-\beta$}{} Parameter Space}\label{subsec:quadrantdef}
To guide the interpretation of CE inspiral through the ejection and dynamical parameters, we define quadrants of the $\xi-\beta$ parameter space that inform us about the hydrodynamics of CE inspiral at a given moment. In particular, we argue that if the energy criterion for ejection is met, the dynamical criterion indicates whether inspiral continues or begins to regulate by effectively coupling to the surrounding envelope.

\begin{enumerate}
    \item In quadrant I (Q1), $\xi>1$ and $\beta>1$. Therefore there is both sufficient energy to unbind the CE ($\xi>1$) and that energy has time to be shared via shock waves before the inspiral continues ($\beta>1$). In this case, we argue that the conditions are met for sufficient ejection, i.e. to eject the local envelope and slow or halt the inspiral.
    \item In quadrant II (Q2), $\xi<1$ and $\beta>1$, which means there is insufficient energy to unbind the envelope ($\xi<1$) but inspiral is slow relative to the sharing of that energy ($\beta>1$). This satisfies the conditions for a comparatively gradual inspiral. This is sometimes described as a self-regulated inspiral \citep{2001ASPC..229..239P,2013A&ARv..21...59I,Yarza2023} because the material surrounding the secondary can be heated by many orbital cycles, raising its entropy and moderating the drag force.   
    \item In quadrant III (Q3), $\xi<1$ and $\beta<1$, which means there is not enough energy to unbind the envelope ($\xi<1$) and the secondary is plunging faster than shocks are propagating ($\beta<1$). This region corresponds to dynamical inspiral in which the envelope is retained, often seen in a merger scenario or prior to a self-regulated phase in successful ejection scenarios.
    \item In quadrant IV (Q4), $\xi>1$ and $\beta<1$. In this case, the deposited orbital energy is sufficient for partial envelope ejection ($\xi>1$), but cannot be shared quickly enough by shocks ($\beta<1$) to cause envelope ejection within $R_\mathrm{a}$ of the secondary. Thus, the secondary's infall continues.
\end{enumerate}

As we will demonstrate, the manner in which a CE inspiral tracks through these quadrants reveals a distinct narrative for that CE event and a prediction for its outcome.

\section{Application} \label{sec:results}

\subsection{Stellar Profiles}\label{subsec:MESA}

In general, we base our model of orbital infall, binding energy, and energy redistribution on the properties of the stellar envelope of the primary star. In order to model representative envelope structures, we use a replicated setup of {\tt MESA} v7503 \citep{Paxton2011,Paxton2013,Paxton2015} containing the inlists and settings of the MIST package \citep{Choi2016,Dotter2016} for primaries with $Z=Z_\odot$ (appropriate for modeling galactic CE events) and, due to limitations in MIST at low metallicity and high mass, {\tt MESA} v10398 \citep{Paxton2011,Paxton2013,Paxton2015,Paxton2018} for primaries with $Z=1/50Z_\odot$ (appropriate for modeling Population III binaries, i.e. progenitors of binary gravitational wave sources)\footnote{The inlists, source files, outputs, and analysis code necessary to reproduce the models and figures in this work are available on Zenodo: \dataset[doi:10.5281/zenodo.14284113]{https://doi.org/10.5281/zenodo.14284113}}.

In what follows, we examine three model systems.  The selected binary combinations represent possible progenitors of the following types, described in terms of  secondary-to-primary mass ratio, $q=M_2/M_1$:
\begin{itemize}
    \item a massive-star merger progenitor, comprised of a $17.8 M_\odot$, $9.5 R_\odot$ primary near TAMS of solar metallicity with secondary $q=0.1$,
    \item a binary neutron star (BNS) progenitor, comprised of a $12.5 M_\odot$ ($M_\mathrm{core}=4.8 M_\odot$), $\approx 800 R_\odot$ primary of solar metallicity with secondary $q=0.15$, and
    \item a binary black hole\footnote{Though there is some evidence to suggest that CE events in high mass binaries contribute to BBH formation \citep[e.g.][]{2017ApJ...845..173M,2021ApJ...910..152Z}, the stability of mass transfer in BBH progenitor systems and its relationship to CE episodes has been called into question \citep{2021ApJ...922..110G}; we include such a system here to gain insight into how CE inspiral would proceed, should it occur, while remaining agnostic as to whether such systems undergo CE events.} (BBH) progenitor, comprised of an $80 M_\odot$ ($M_\mathrm{core}=42.6 M_\odot$), $\approx 1200 R_\odot$ primary  of metallicity $1/50 Z_\odot$ with secondary $q=0.4$. 
\end{itemize} 
The merger and BNS profiles match the initial conditions for 3D hydrodynamical CE simulations from \citet{Tenley} and \citet{Jamie}, respectively, in order to provide a direct comparison which we discuss in Section \ref{subsec:FLASH}.

\subsection{Binary Profiles in Terms of the Ejection and Dynamical Parameters} \label{subsec:quadrantMESA}

\begin{figure*}[tbp]
    \epsscale{1.15}
    \plotone{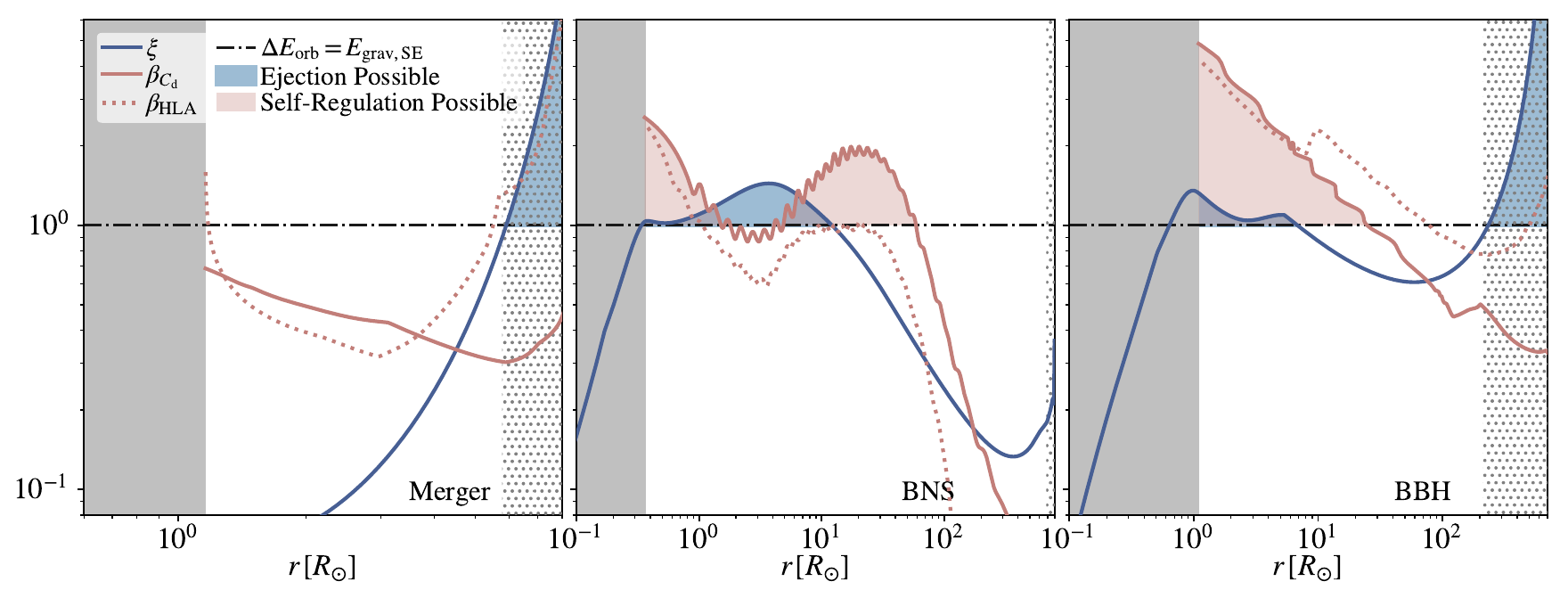}
    \caption{
    Representative examples of ejection parameter $\xi$ and $\beta$ in three types of CE systems. Dark blue curves denote the change in orbital energy \Eorb\ with respect to $r$ normalized by the local value of modified gravitational binding energy $E_\mathrm{grav,SE}$ (Equation~\ref{eqn:Egrav+}). Shaded blue regions represent where the energy criterion for ejection of the stellar envelope outside of location $r-R_\mathrm{a}$ is met. Rose curves denote the ratio of the inspiral timescale to the shock-heating timescale integrated using the HLA (dotted) and CE drag formalism (solid) prescriptions, with shaded pink regions representing where the hydrodynamical criterion for sufficient ejection is met according to the CE drag formalism. The core of the primary is shown in grey as defined by hydrogen mass fraction $X_\mathrm{H} < 0.1$, with expected pre-CE mass loss denoted by the dotted outer region. \emph{Left}: A merger progenitor system with TAMS primary of $17.8M_\odot$ and mass ratio $0.1$, as used for the merger simulations of \citet{Tenley} and \citet{2024ApJ...971..132E}. \emph{Center}: A BNS progenitor system with a $12.5 M_\odot$ red supergiant with mass ratio $0.15$, as used for the BNS progenitor CE simulations of \citet{Jamie}. \emph{Right}: A BBH progenitor system with an AGB primary of $80 M_\odot$ and mass ratio $0.4$.} \label{fig:combined}
\end{figure*}

To explore CE inspiral and possible ejection in a given system, we now illustrate how $\xi$ and $\beta$ co-evolve for our selection of representative CE binaries (Section \ref{subsec:MESA}). 
Figure \ref{fig:combined} shows the values of both dimensionless parameters as a function of radius within our model systems. We shade regions where either $\beta>1$ or $\xi>1$, emphasizing that these criteria are not linked: they do not necessarily occur in the same regions of the envelope.  The core of each primary is excluded  (grey shading), as is the outermost envelope (dotted shading), which is lost in the mass transfer that leads-up to the CE encounter, in accordance with the pre-CE mass loss prescription of \citet{MacLeod2020b}. 
CE simulations consistently show that onset leading to plunge takes place in these outer regions, even when the energy criterion ($\xi>1$) is locally satisfied \citep[e.g., ][]{2019MNRAS.490.3727C,2019MNRAS.484..631R,2021MNRAS.500.1921G,Tenley,2024ApJ...971..132E}. 

To determine inspiral rates, we adopt two alternatives. We apply the drag prescriptions from HLA and the CE drag formalism, which is equivalent to HLA but with a drag coefficient motivated by numerical calculations (see Appendix \ref{app:CEdrag}), to calculate $\beta_\mathrm{HLA}$ and $\beta_{C_\mathrm{d}}$, respectively. 
In either case, we integrate the orbital motion of the secondary through the primary's envelope to derive the inspiral timescale of  Equation \ref{eqn:tau_insp} and the instantaneous orbital velocity, $v_\infty$, which we use to define $\tau_\mathrm{SH}=\pi r/v_\infty$ from Equation \ref{eqn:tau_SH}, giving an upper limit for $\beta$. Though neither prescription is integral to our model, and could be replaced by an alternative derivation of  $\beta$, we compare them to illustrate this framework's ability to be calibrated to better reflect the behavior seen in 3D simulations.\footnote{We note here that $\xi$ values are not impacted by the choice of drag prescription, and that not all CE binaries will show drastic differences in $\beta$ between the drag prescriptions, especially those with relatively shallow density gradients in the envelope \citep{De2019}.}
 For example, in the BBH case (right panel, Figure \ref{fig:combined}), the $\beta_\mathrm{HLA}$ curve lies above unity through $\approx 60\%$ of the stellar envelope (by radius), suggesting a regulated inspiral. By contrast, the $\beta_{C_\mathrm{d}}$ curve lies below unity in all but a few percent of the inner envelope, suggesting a steep plunge that is not energy-regulated.

\begin{figure}[tbp]
    \epsscale{1.1}
    \plotone{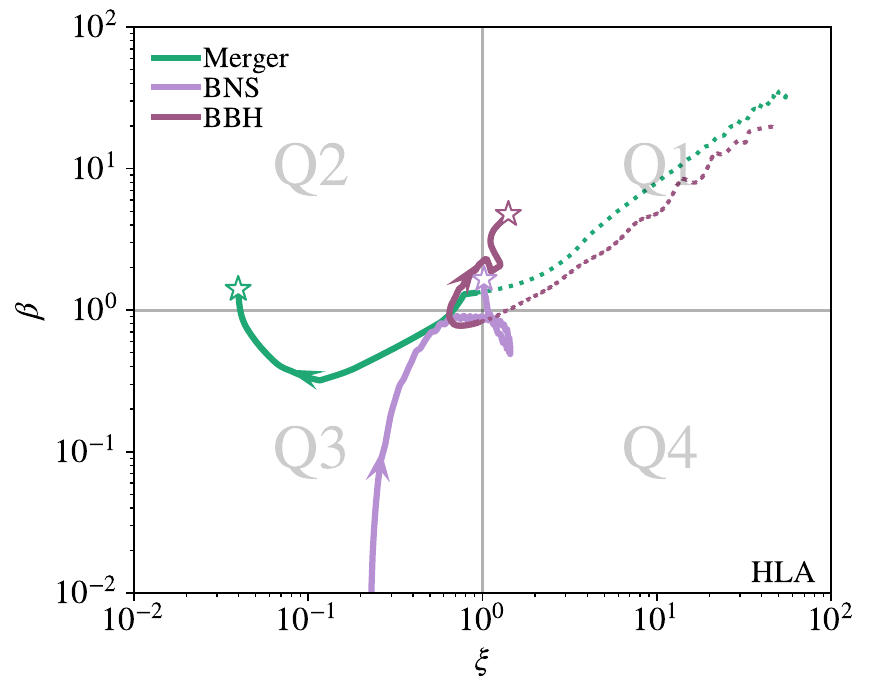}
    \plotone{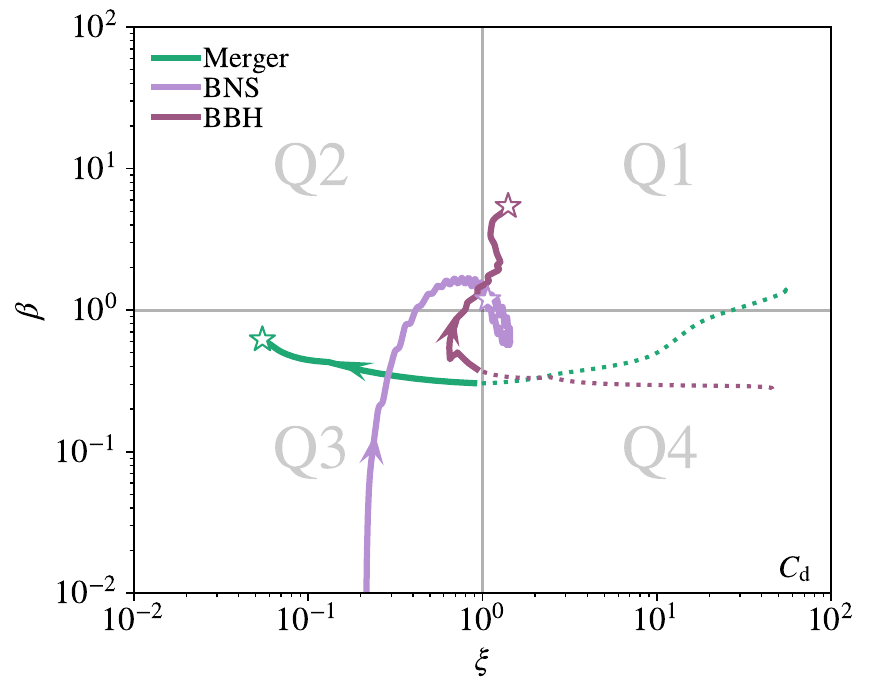}
    \caption{Integrated results for the HLA (top) and CE drag formalism (bottom) prescriptions in the $\xi-\beta$ parameter space for three representative example CE binaries: a merger progenitor of a TAMS $17.8 M_\odot$ star with secondary $q=0.1$ (green), a BNS progenitor of $12.5 M_\odot$ with secondary $q=0.15$ (purple), and a BBH progenitor of $80 M_\odot$ and secondary $q=0.4$ (dark plum). Dotted lines denote pre-CE mass loss, solid lines denote integrated CE inspiral, and stars denote the core boundary where integration was stopped. } \label{fig:quad}
\end{figure}

Figure \ref{fig:quad} moves to view a CE inspiral in terms of the $\xi-\beta$ parameter space defined in Section \ref{subsec:quadrantdef}. In this context, individual episodes are represented by tracks through the $\xi-\beta$ space. Dashed lines denote the outermost envelope (as in dot-shaded regions of Figure \ref{fig:combined}) and stars represent the core boundary of the stellar profile. 
As in Figure \ref{fig:combined}, we compare predictions from HLA and numerical $C_{\rm d}$ in the upper and lower panels of Figure \ref{fig:quad}. 
As anticipated from Figure \ref{fig:combined}, tracks in the HLA panel tend toward higher $\beta$ values, which is expected as the HLA drag prescription tends to underestimate drag in CEs, leading to shallower inspiral overall \citep{2015ApJ...798L..19M,2015ApJ...803...41M,2017ApJ...838...56M}. 

The curves in the CE drag formalism panel show several differences that are more consistent with 3D models (Section \ref{subsec:FLASH}). Onset (dotted) passes through Q4 (bottom) rather than Q1 (top), which indicates the type of plunge concurrent with mass ejection that is often seen during onset in 3D hydrodynamical models rather than sufficient ejection at wide separation. 
In the merger case (green), HLA (top) predicts a slow onset (Q1) and self-regulated merger (Q2), while numerical $C_\mathrm{d}$ (bottom) predicts a steep plunge that ejects the outer envelope without slowing inspiral (Q4) and a dynamical merger (Q3), again consistent with 3D models (see Section \ref{subsec:FLASH}). The HLA BNS progenitor system (purple, top) begins with a very steep Q3 plunge that slows, then steepens slightly in Q4 before looping back to eject the envelope in Q1 very close to the core. This implies a dynamical ejection, or envelope ejection directly from the dynamical inspiral stage of CE. Though we see similar behavior for numerical $C_\mathrm{d}$ (bottom), it includes a shift into the self-regulated (Q2) and sufficient ejection (Q1) regions prior to the loop. This implies that ejection would likely occur on the first pass, rather than the second, at a larger separation and after a typical self-regulated phase. If we exclude onset in the BBH case (dark plum) for both HLA and $C_\mathrm{d}$, a dynamical plunge (Q3) is slowed as it enters material that is more efficiently shock-heated (Q2), leading to successful ejection (Q1) within $10 R_\odot$ of the core after a self-regulated inspiral. This is the closest representation to the typical narrative of a successful CE event, which is expected to include a dynamical inspiral, self-regulated phase, and envelope ejection, in that order.

The BBH and merger cases in particular have implications for the energy formalism: these systems, according to the standard energy formalism, would have an orbital energy budget capable of ejecting the envelope at very wide separation to the point of avoiding a CE episode altogether. Yet with these early stages lying in (or near) Q4, their $\beta$ profiles suggest that a dynamical inspiral may be favored even with an energy surplus, allowing a degree of tightening that, in the BBH case, would reduce the system's separation by nearly two orders of magnitude. This is a remarkable difference in predicted outcome.

\subsection{Two Case Studies in 3D Hydrodynamics} \label{subsec:FLASH}
In what follows, we apply the $\xi-\beta$ framework to two detailed 3D hydrodynamical simulations to assess  whether this model is consistent with simulation outcomes. We apply a customized setup using the 3D adaptive-mesh refinement hydrodynamics code {\tt FLASH} v4.3 \citep{2000ApJS..131..273F} that is able to load and fully resolve {\tt MESA} stellar profiles in 3D with a Helmholtz equation of state for the simulation of CE interactions \citep{2009ApJ...705..844G,2013ApJ...767...25G,Jamie,2020ApJ...901...44W,Tenley}. 
We adapt the models of \citet{Tenley} and \citet{Jamie} to simulate a merger and a BNS progenitor scenario, respectively (for further details, see Appendix \ref{app:FLASHsetup}). The model of \citet{Jamie} is based on an extended stellar profile with the outer envelope (assumed ejected in accordance with the energy formalism) trimmed to $10 R_\odot$ to achieve core resolution, initiating with inspiral already underway at $8 R_\odot$, therefore $\xi$ and $\beta$ quantities are calculated based on the initial untrimmed stellar profile in the unsimulated regions. In both cases, initial conditions for the dynamics of the secondary are calculated to match those derived from the CE drag formalism inspiral integration for the same systems in Section \ref{subsec:quadrantMESA}.

\begin{figure}[tbp]
    \epsscale{1.1}
    \plotone{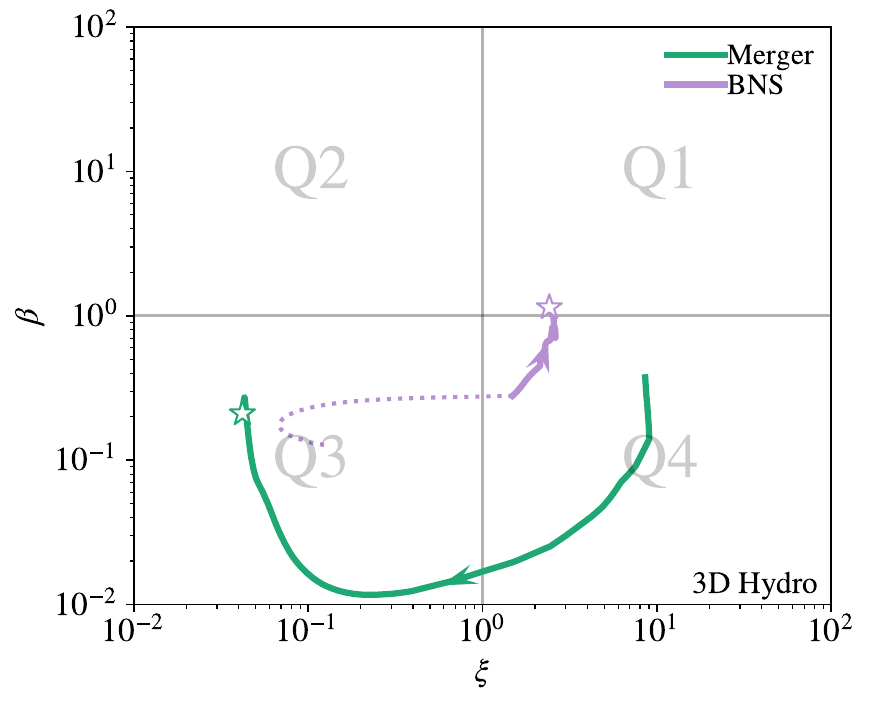}
    \caption{Simulation results from 3D hydrodynamical models of two CE binaries in the $\xi-\beta$ parameter space. A high-resolution core-resolved model of the merger of a $1.57 M_\odot$ neutron star with a TAMS $17.8 M_\odot$ primary is shown in green, based on the setup of \citet{Tenley} and \citet{2024ApJ...971..132E}. A high-resolution core-resolved model of a successful CE ejection with a $1.4 M_\odot$ neutron star and a $12 M_\odot$ red supergiant stripped to $10 R_\odot$ is shown in purple, based on the setup of \citet{Jamie}. Solid lines denote calculations from simulated inspiral, dotted lines denote calculations from integrated inspiral in non-simulated regions, and stars denote the final position of the secondary: central in the case of the merger, and circularized at $\approx 2 R_\odot$in the case of the BNS progenitor.} \label{fig:quadsim}
\end{figure}

The simulation results in the $\xi-\beta$ parameter space are shown in Figure \ref{fig:quadsim}. The dynamical inspiral in the merger simulation (green), though notably lower in $\beta$ value, is qualitatively similar to that seen in the CE drag formalism inspiral integration in Figure \ref{fig:quad}. As the curve proceeds right to left, a steep plunge ejects $\approx 10\%$ of the envelope by mass without interrupting the inspiral, then begins to relax before the secondary settles into the core. Notably, if we consider the energy criterion alone, as we do with the standard energy formalism, this system would be expected to undergo envelope ejection in the outer region in which $\xi > 1$ and remain a binary, yet this does not occur here. It is not only possible, but seen across the board in 3D hydrodynamical CE merger models \citep[e.g.,][]{2021MNRAS.500.1921G,Tenley,2024ApJ...971..132E}, that some envelope may be ejected as inspiral continues; this is easily explained when we note that $\beta < 1$ in this region, implying that sufficient ejection is not hydrodynamically favorable.

Though less obvious, the BNS progenitor simulation (purple) also shows a qualitative similarity to that seen in the inspiral integration, but only the final stage leading to sufficient ejection. In Figure \ref{fig:quad}, there is a Q4 inflection point in the $\xi-\beta$ trajectory of the BNS progenitor system in which the inspiral quickly reverses from becoming more dynamical to becoming less dynamical: this inflection point occurs at the maximum $\xi$ value after the inspiral has already tracked through the sufficient ejection region (Q1) once, and the starting radius of inspiral in the simulation is just outside this peak.

We emphasize here that it is unlikely for an inspiral to proceed through Q1 into another quadrant, as we suggest that inspiral will terminate when both energy and hydrodynamical conditions are favorable for sufficient ejection. Notably, this simulation does not pass through the outer self-regulated region (Q2, $\approx 7-15 R_\odot$ for this primary), but initiates just inside of it at $8 R_\odot$ with envelope beyond $10 R_\odot$ already removed, leading to envelope ejection and circularization at $\approx 2 R_\odot$. That outcome is consistent with what we see in Figure \ref{fig:quad} if we begin tracking inspiral in Q4 near the inflection point. However, the complete $\beta$ profile in Figure \ref{fig:quad} (lower panel) suggests that if this setup were to initiate inspiral at a distance beyond the outer self-regulated region, sufficient ejection might occur at the first pass through Q1 at a distance of $\approx 10 R_\odot$, interrupting further tightening.

However, with inspiral initiated \textit{in medias res} in Q4, both the integrated curves and the simulated curves increase in $\beta$ value until they reach the sufficient ejection region (Q1) next to the core. The simulation ends with approximately twice the $\xi$ value of the HLA and CE drag formalism inspirals, which would likely be resolved with an improved method for calculating $E_\mathrm{grav,SE}$ near the primary's core in 1D (see Section \ref{subsec:caveats}).

\section{Discussion} \label{sec:discussion}

It has long been the goal of the CE community to develop a prescription that can approximate the results of the best state-of-the-art 3D models with the fewest inputs and at the smallest computational expense possible.
Following  previous efforts to improve upon the widely-utilized standard energy formalism \citep[e.g.,][]{deKool1990,2011ApJ...731L..36I,2019MNRAS.485.4492W,2022ApJ...937L..42H,2022MNRAS.516.2189W,2023ApJ...944...87D,2024arXiv240604118N}, this framework is the first to implement a component predictive of CE hydrodynamic wave speeds into an analytical model.

\subsection{Additional Considerations} \label{subsec:caveats}

While we have considered the hydrodynamic aspects of CE phases, there are a number of processes that require further consideration. In particular, while our model is predictive of the hydrodynamic behavior, it is not yet clear whether it is fully predictive of the CE outcome. Yet, we believe it can be naturally extended. For example, we have so far considered the gravitational binding energy and excluded other energy sources such as internal or recombination energy in our calculation of the energetic parameter, $\xi$. These might be included in an extended model. 

At higher $q$ values, the companion's gravitational reach may be better described by the Roche radius \citep{1983ApJ...268..368E} than $R_\mathrm{a}$. Such considerations are relevant in higher $q$ binaries, as well as late stages of CE when the secondary nears the primary's core, as the assumptions behind the drag formalism break down when $R_\mathrm{a}$ penetrates the core. Therefore, the calculation for $\xi$ would benefit from a more sophisticated treatment for such cases. In fact, CE wind tunnel simulations \citep{2015ApJ...798L..19M,2017ApJ...838...56M,De2019} 
 show that the presence of a steep density gradient sharply redirects shock-heated material outward as waves couple more easily with less dense material, suggesting that the core may be preserved and no part of its binding energy need be included in $E_\mathrm{grav, SE}$, giving higher $\xi$ values in this region.

Further, we suggest that additional consideration will need to be included to best describe the ``stopping'' criteria of events. For example, the CE systems detailed in this work generally have primary cores massive enough to avoid disruption by the secondary during inspiral, but this is not always the case. Some treatment for the tidal disruption of either the primary or the secondary should be included to enhance the framework's predictive capabilities \citep[][]{2002MNRAS.334..819I}. Secondly, when a CE achieves the ejection criteria of $\xi>1$ and $\beta>1$, it may leave behind some of the H-rich envelope interior to the current separation. This is not necessarily problematic: if the remaining envelope  contracts onto the core any ongoing interaction will ``shut off.'' On the other hand, if the remaining envelope expands as the overlying layers are removed it would drive continued interaction \citep{1997A&A...327..620S,2011ApJ...730...76I,2016MNRAS.462..362I,2020PASA...37...38V}.

\subsection{Summary and Conclusion}
In this work, we reexamined the standard CE energy formalism through the lens of inspiral hydrodynamics, introducing adjustments and additions that together create a new CE analytical framework for predicting envelope ejection sufficient to end inspiral. The $\xi-\beta$ framework combines two essential controlling parameters of the CE process: the ejection parameter, $\xi$, which defines where and when envelope ejection is energetically favorable, and the dynamical parameter, $\beta$, which defines where and when envelope ejection is hydrodynamically favorable. 

To illustrate, we presented 1D analyses of a selection of CE binaries, applying drag prescriptions from HLA and the CE drag formalism, and performed two 3D hydrodynamical case studies which we analyzed with the $\xi-\beta$ framework, finding that the simulation results are broadly consistent with the integrated results based on the initial, 1D profiles. 
Further work is needed for refinement and validation more broadly, but in its current form, the $\xi-\beta$ framework provides a straightforward method for interpreting and predicting how inspiral will proceed in a CE episode.

Notably, this framework suggests that energy transport that is slow relative to infall causes inspiral to continue in spite of an energy excess. This encourages further investigation of CE evolution as a potential formation channel for BBHs, as well as other ``excess energy'' systems that would nominally end CE phases at wide separations in previous analyses.

\begin{acknowledgments}
    We thank D. Lin, M. Gallegos-Garcia, N. Soker, J. Law-Smith, A. Loeb, S. Toonen, and R. Yarza for helpful discussions. We also gratefully acknowledge the anonymous referee for their insightful feedback. R.W.E. acknowledges the support of the University of California President's Dissertation-Year Fellowship, the Heising-Simons Foundation, and the Vera Rubin Presidential Chair for Diversity at UCSC. M.M. gratefully acknowledges support from a Clay Fellowship at the Smithsonian Astrophysical Observatory. E.R.-R. acknowledges support from the Heising-Simons Foundation and NSF (AST-2307710, AST-2206243, AST-1911206, and AST-1852393). Any opinions, findings, and conclusions or recommendations expressed in this material are those of the authors and do not necessarily reflect the views of the NSF. This work was supported in part by NASA ATP Grant 80NSSC22K0722 to the University of North Carolina at Chapel Hill. The 3D hydrodynamics software used in this work was developed in part by the DOE NNSA- and DOE Office of Science-supported Flash Center for Computational Science at the University of Chicago and the University of Rochester. We acknowledge use of the computational resources of the SCIENCE HPC Center at the University of Copenhagen, with the support of the DARK Cosmology Centre at the Niels Bohr Institute.
\end{acknowledgments}

\vspace{5mm}
\software{
{\tt MESA} \citep{Paxton2011, Paxton2013, Paxton2015, Paxton2018}, 
{\tt FLASH} \citep{2000ApJS..131..273F},
          {\tt matplotlib} \citep{Hunter2007}, 
          {\tt NumPy} \citep{vanderwalt2011}, 
          {\tt py\_mesa\_reader} \citep{WolfSchwab2017}}


\appendix
\section{The CE Drag Formalism} \label{app:CEdrag}
A typical CE binary, in which the primary is more massive than the secondary, is characterized by the global mass ratio $q=M_2/M_1<1$. Inspiral and envelope ejection, however, are partially characterized by the radius-dependent local mass ratio
\begin{equation} \label{eqn:qr}
    q_\mathrm{r} = \frac{M_2}{M_\mathrm{enc}} ,
\end{equation}
which increases as the orbit decays and can approach unity in high mass, high mass ratio ($q>0.4$) systems as the secondary nears the core.

The velocity of the secondary during inspiral can be described as a modified Keplerian orbit:
\begin{equation} \label{eqn:vkep}
v_\infty = f v_\mathrm{Kep} = f \sqrt{\frac{G (M_\mathrm{enc} + M_\mathrm{2})}{r} } ,
\end{equation}
in which $f$ is the fraction of Keplerian velocity of the embedded object relative to the envelope. In the $f$ term is incorporated both the degree of corotation with the envelope and the degree to which drag has reduced the velocity from Keplerian.

The relative velocity $v_\infty$, when combined with the stellar structure of the primary and the mass ratio of the system, defines several key quantities relevant to how inspiral proceeds, borrowing from the framework for flows and accretion developed by \citet{Hoyle1939} and \citet{Bondi1944} (HLA). The movement of the secondary relative to the flow of the envelope is characterized by the Mach number
\begin{equation}
    \mathcal{M}_\infty = \frac{v_\infty}{c_\mathrm{s}},
\end{equation}
in which $c_\mathrm{s}$ is the sound speed of the oncoming gas. Except in cases of very low mass ratio, typical CE inspirals span a range of $\mathcal{M}\sim 1-5$, with higher values possible in the outer envelope, though these regions are often disturbed or expelled due to mass transfer before CE onset \citep{MacLeod2020b}.

In HLA and the CE drag formalism, the characteristic length scale for the gravitational reach of the secondary on the surrounding gas is the accretion radius
\begin{equation} \label{eqn:Raappendix}
R_\mathrm{a} = \frac{2 G M_\mathrm{2}}{v_{\infty}^2},
\end{equation}
effectively a gravitational ``sphere of influence.'' Envelope material within this radius will be attracted and produce a drag force.

For envelopes that can be approximated as a constant entropy polytrope, the dimensionless density gradient $\epsilon_\rho$ is directly related to the Mach number and mass ratio by the following relation shown by \citet{2020ApJ...899...77E}:
\begin{equation} \label{eqn:erhorelation}
    \epsilon_\rho = \frac{2 q_\mathrm{r}}{(1+q_\mathrm{r})^2} \mathcal{M}^2_\infty ,
\end{equation}
meaning only two of these flow parameters must be known in order to characterize the effects of the flow on the CE inspiral.

These effects are quantified in the CE drag formalism by the drag coefficient $C_\mathrm{d}$, the value of which derives from the steady-state time-averaged drag force $F_\mathrm{d}$ measured in the suite of CE wind tunnel simulations normalized by the HLA drag force such that
\begin{equation} \label{eqn:Fdrag}
    F_\mathrm{d} = C_\mathrm{d} \pi R_\mathrm{a}^2
    \rho_\infty v_\infty^2 ,
\end{equation}
in which $\rho_\infty$ is the local density at radius $r$. The functional forms of $C_\mathrm{d}$ for $\gamma=4/3$ and $5/3$ polytropic equations of state were introduced by \citet{De2019}, allowing the straightforward integration of the drag force for inspirals through a broad range of stellar profiles, and in turn the orbital energy deposition rate through the relation
\begin{equation} \label{eqn:edot}
    \dot{E}_\mathrm{orb} =
    -F_\mathrm{d} v_\infty .
\end{equation}

With Equations \ref{eqn:Fdrag} and \ref{eqn:edot}, the CE drag formalism provides two useful quantities to approximate hydrodynamical phenomena in the 1D analysis of a CE event: a proxy for the reach of energy deposition during inspiral (Section \ref{subsec:ejectionparam}) and a timeline for energy deposition and inspiral (Section \ref{subsec:dynamicalparam}).

\section{3D Hydrodynamical Setup} \label{app:FLASHsetup}
In Section \ref{subsec:FLASH}, we perform two high-resolution 3D hydrodynamical simulations using a customized setup of the 3D adaptive-mesh refinement hydrodynamics code {\tt FLASH} v4.3 \citep{2000ApJS..131..273F}. The specific setups used in this work are based on those of \citet{Jamie} and \citet{Tenley}, which in turn were based on that of \citet{2020ApJ...901...44W}, which was based on those developed by \citet{2009ApJ...705..844G} and \citet{2013ApJ...767...25G}. This setup utilizes a Helmholtz equation of state in an extended Helmholtz table which assumes full ionization \citep{2000ApJS..126..501T}, therefore does not include recombination energy. The 1D profile of a stellar model generated with {\tt MESA} is loaded into the setup and relaxed onto the grid for approximately $100$ core dynamical times, after which the secondary is initialized as a point mass with gradually increasing mass and velocity until it matches the initial conditions defined for the simulation.

In the merger case, we use a domain size of $\Delta X_\mathrm{max}=60 R_\odot$ on each side. We apply a resolution of $\Delta X_\mathrm{min}\approx 0.029 R_\odot$ to improve fidelity around the stellar core, and adjust initial conditions to reflect those of the CE drag formalism inspiral integration with inspiral beginning at the limb of the primary with secondary at Keplerian velocity. The full stellar model of the primary is included in the simulation.

In the BNS progenitor case, we use a domain size of $\Delta X_\mathrm{max}=40 R_\odot$ on each side. We apply a resolution of $\Delta X_\mathrm{min}\approx 0.0098 R_\odot$ and trim the envelope to $10 R_\odot$ before loading onto the grid. We initialize the secondary at $8 R_\odot$ as described by \citet{Jamie}, but again adjust inspiral initial conditions to reflect those of the CE drag formalism inspiral integration.

For a comprehensive review of the development and evolution of these setups, see \citet{2009ApJ...705..844G,2013ApJ...767...25G,2020ApJ...901...44W,Jamie,Tenley}.

\bibliographystyle{aasjournal}
\bibliography{refs}{}

\end{document}